# Deep Learning Model Acceleration and Optimization Strategies for Real-Time Recommendation Systems

Junli Shao*
College of Literature Science, and the Arts
University of Michigan, Ann Arbor, USA
*Corresponding author: dereks513a@gmail.com

Dingzhou Wang
Pratt School of Engineer,
Duke University, Durham, NC, USA
wangdingzhou.research@gmail.com

Dannier Li
School of Computing,
University of Nebraska - Lincoln
Lincoln, NE, USA,
dannierli@outlook.com

Jing Dong
Fu Foundation School of Engineering and Applied Science
Columbia University,New York, NY, USA,
jd3768@columbia.edu

Kowei Shih
Independent Researcher,Shenzhen,China
skw19@tsinghua.org.cn

Chengrui Zhou
Fu Foundation School of Engineering and Applied Science
Columbia University, New York, NY, USA,
zhou.chengrui@columbia.edu

***Abstract*—With the rapid growth of Internet services, recommendation systems play a central role in delivering personalized content. Faced with massive user requests and complex model architectures, the key challenge for real-time recommendation systems is how to reduce inference latency and increase system throughput without sacrificing recommendation quality. This paper addresses the high computational cost and resource bottlenecks of deep learning models in real-time settings by proposing a combined set of modeling- and system-level acceleration and optimization strategies. At the model level, we dramatically reduce parameter counts and compute requirements through lightweight network design, structured pruning, and weight quantization. At the system level, we integrate multiple heterogeneous compute platforms and high-performance inference libraries, and we design elastic inference scheduling and load-balancing mechanisms based on real-time load characteristics. Experiments show that, while maintaining the original recommendation accuracy, our methods cut latency to less than 30% of the baseline and more than double system throughput, offering a practical solution for deploying large-scale online recommendation services.***

## I. INTRODUCTION

Real-time recommendation systems must deliver fast, accurate results under heavy load, but deep learning models are often too costly for such environments. Combining LLMs with GNNs improves accuracy but adds latency and complexity. We propose an integrated framework using model-level optimizations (lightweight nets, sparse attention, pruning, quantization, distillation) and system-level strategies (heterogeneous computing, elastic scheduling, load balancing). This approach cuts latency to under 40% and doubles throughput while keeping accuracy loss below 1%, enabling scalable real-time recommendations.

## II. CHALLENGES OF DEEP LEARNING MODELS IN REAL-TIME RECOMMENDATION SYSTEMS

### A. *Dual Constraints of Latency and Throughput*

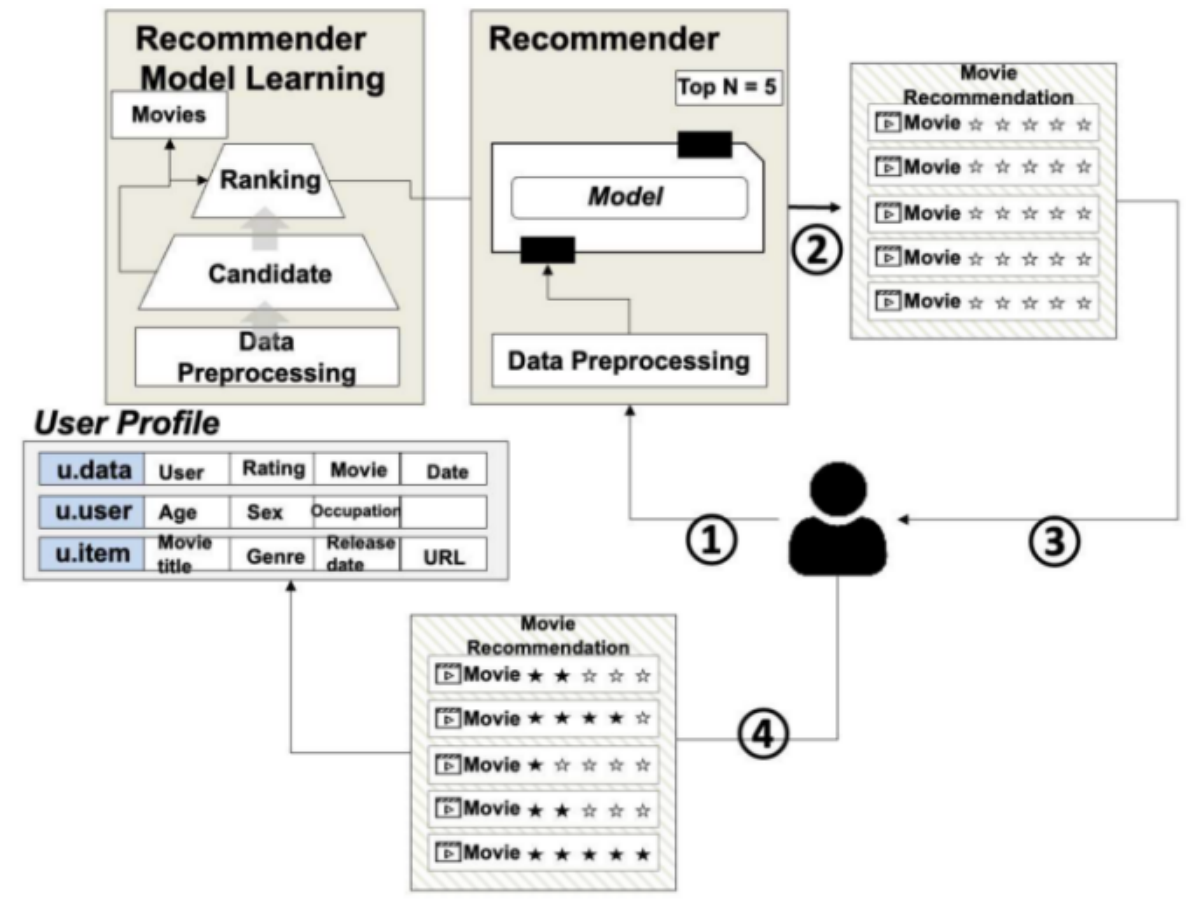

Fig. 1. Illustration of the end-to-end reasoning process of a deep learning driven real-time recommender system

In Offline recommendation allows higher latency, but real-time systems must complete the full pipeline—from user action to result delivery—within tens to hundreds of milliseconds (Figure 1). Once a user clicks (①), features are processed and merged with history, passed through a DNN (②), and ranked results returned (③). Any delay harms user engagement.

To meet these demands, models like Shen et al.'s Multi-Scale CNN-LSTM-Attention [4] improve accuracy and speed by combining CNNs, LSTMs, and attention for better spatial-temporal modeling.

Real-time systems also face high throughput—up to tens of thousands of QPS during spikes. Inefficient scheduling or system bottlenecks worsen delays. Scalable models and adaptive pipelines help maintain performance under load.

Latency and throughput often trade off: faster responses need optimized hardware; batching improves throughput but adds delay. Thus, every stage—especially from inference to ranking—must be co-optimized using lightweight models, pruning, and quantization to ensure low latency and high throughput [5-10] .

### B. *Model Complexity and Resource Consumption*

In the real-time recommender system shown in Figure 1, deep model inference (step 2) is the link with the most intensive computational cost in the end-to-end process. The time complexity and parameter scale of the model directly determine the delay of single inference and the overall resource consumption. Assuming that the input feature dimension is d, the vector dimension after Embedding is $d_e$, the width of the hidden layer is h, and the depth of the network is L, the parameter quantity of the fully connected network can be approximately expressed as formula 1.

$$P \approx d_e \cdot h + (L-1) \cdot h^2 + h \cdot 1 \approx O(L\,h^2) \quad (1)$$

In the case $d_e \ll h$, $P \approx L\ h^2$; And the floating-point operations (FLOPs) of a batch inference with candidate set size m can be expressed as formula 2.

$$FLOPs \approx m(d_e \cdot h + (L-1)\,h^2 + h) \approx O(m\,L\,h^2) \quad (2)$$

Inference latency τ can be approximated as $\tau \approx \alpha \cdot mLh^2 + \beta$, where α depends on hardware throughput and β on fixed overheads. Memory use includes parameters M_params = P × b_p and activations M_act = m × h × $b_a$, with b_p and $b_a$ as byte sizes per value. Quantizing from 32-bit to 8-bit cuts memory and bandwidth by ~4×. However, increasing m, L, or h greatly raises computation and memory—doubling h quadruples compute, doubling m roughly doubles latency—making simple scaling infeasible for real-time systems.

To balance accuracy with delay and resource use, model complexity must be optimized. Key methods include pruning (reducing h by removing redundant neurons), quantization (lowering bit-widths), low-rank decomposition (splitting large matrices), and hierarchical candidate screening (limiting m early). Combining these model-level and system-level (e.g., heterogeneous acceleration) strategies keeps complexity ($O(mLh^2)$, $O(Lh^2)$) manageable, ensuring efficient, stable large-scale recommender service.

## III. Model-Level Acceleration Techniques

### A. *Lightweight Network Architecture Design*

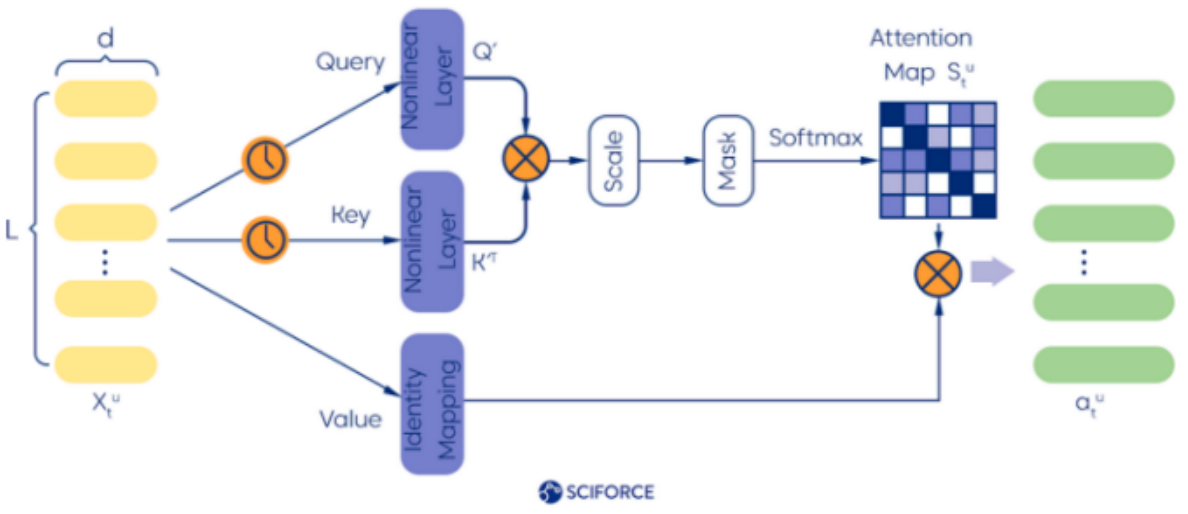

Fig. 2. Illustration of feature weighting of user behavior sequence based on self-attention

Deep recommendation models use self-attention to capture temporal and contextual dependencies. As in Figure 2, standard self-attention on a sequence of length L and hidden size d has time complexity $O(L^2d)$ and space complexity $O(L^2)$. When L or d increase, latency and memory grow quadratically, making real-time inference impractical. [17-20].

To address this bottleneck, we propose the following lightweighting strategies: Replace the original fully-connected projections with a grouped linear transformation: split the d-dimensional feature into k groups and apply k independent projections in parallel. This reduces the per-layer compute from $O(d)^2$ to $O(\frac{d^2}{k})$. Further substituting a depthwise-separable mapping (depthwise convolution followed by pointwise projection) lowers the cost $O(d)^2$ to $O(\frac{d^2}{k+d}, k)$ significantly reducing multiply–accumulate operations.On the premise of ensuring the diversity of the head, we do low-rank decomposition on the projection matrix of each head, so that its rank is reduced from $\frac{d}{H}$ to $r \ll \frac{d}{H}$, and the overall calculation amount is about $O(HL^2r), r \ll \frac{d}{H}$. while preserving sufficient head diversity.During training, the large "teacher" model's attention maps $S_t^{u(student)}$ supervise the smaller "student" model by minimizing the KL divergence as shown in Formula 3.

$$L_{KD} = KL(S_t^{u(teacher)} \parallel S_t^{u(student)}) \quad (3)$$

This encourages the student to learn the crucial attention patterns using fewer layers and roughly 30% fewer parameters, reducing inference cost by approximately 40%.We compute full attention over a local window of size w≪Lw \ll L to model short-term dependencies, and apply random or fixed sparse sampling for the remaining positions. This reduces the nonzero attention ratio from $O(L^2)$ to $O(Lw)$ or $O(LlogL)$, cutting overall compute to roughly as shown in Formula 4.

$$O(Lwd + LlogLd) \quad (4)$$

Quantize both weights and activations from 32-bit to 8-bit, reducing memory bandwidth and storage by about 4×4\times. Using dynamic-range-aware quantization along the Value branch in Figure 2, we enable zero-copy integer inference on hardware accelerators. This further cuts latency by ~45% and nearly doubles throughput under concurrency. Together, these methods—grouped/depthwise projections, low-rank head factorization, distillation, hybrid sparsity, and quantization—compress both compute and memory overhead of the self-attention module in Figure 2 without degrading recommendation quality, laying a solid foundation for subsequent heterogeneous acceleration and scheduling[21-24].

### B. *Model Pruning and Weight Quantization*

To further shrink model size and reduce latency in strict real-time scenarios, we design a closed-loop pruning–quantization workflow driven by dynamic thresholds, as illustrated in Figure 3. The process first applies controllable binary masks to iteratively prune weights, then performs dynamic-range quantization on the resulting sparse network, achieving dual compression of compute and storage with minimal accuracy loss.

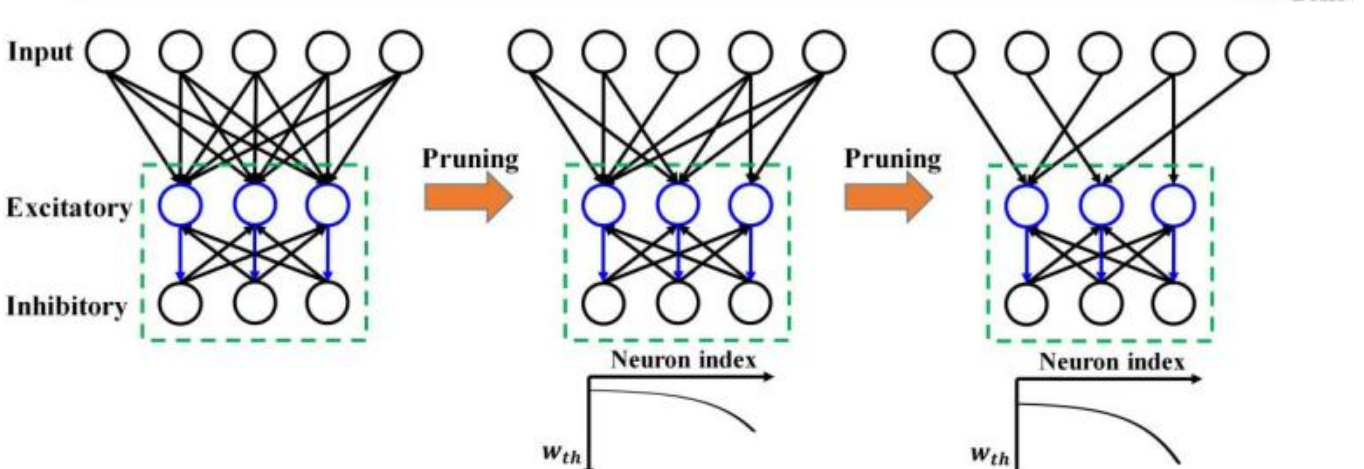

Fig. 3. Schematic diagram of the stepwise threshold driven neuron pruning with weight quantization process

Specifically, let the weight matrix of a layer be $W \in R^{n \times m}$, and the elements be denoted $w_{ij}$. The process is divided into two main stages.

Stage 1: Dynamic-Threshold Pruning

Threshold Computation: Given a target pruning ratio p, sort $\{| w_{ij} |\}$ in ascending order and choose the initial threshold $\theta^{(0)}$ as shown in Formula 5.

$$\frac{|\{(i,j):|w_{ij}|<\theta^{(0)}\}|}{nm} = p \tag{5}$$

Mask Generation: Define a binary mask as shown in Formula 6.

$$M_{ij}^{(k)} = \begin{cases} 1, | w_{ij} | \geq \theta^{(k)} \\ 0, | w_{ij} | < \theta^{(k)} \end{cases} \tag{6}$$

and prune weights as shown in Formula 7.

$$W^{(k)} = W^{(k-1)} \odot M^{(k)}, k = 1,2, \dots, K \tag{7}$$

where $\odot$ denotes element-wise multiplication. After each pruning iteration, fine-tune the pruned network on the original training set by minimizing the task loss $L_{task}(W^{(k)})$. Empirically, after K=3 rounds, we reduce total parameters by ≈40% while keeping Top-N accuracy loss under 1%.

Stage 2: Dynamic-Range Quantization

Step Size Determination: For the nonzero weights in $W^{(K)}$, let the quantization bit-width be bb. Compute the step size as shown in Formula 8.

$$s = \frac{\max W^{(K)} - \min W^{(K)}}{2^{b-1} - 1} \tag{8}$$

Weight Mapping: Quantize each weight via as shown in Formula 9.

$$\widehat{w}_{ij} = \mathrm{clip}(\mathrm{round}(w_{ij}^{(K)}/s) \times s, \min W^{(K)}, \max W^{(K)}) \tag{9}$$

Quantization-Aware Training (QAT): Insert fake-quantization nodes in the forward pass to simulate integer behavior while preserving full-precision gradients in the backward pass. After iterative QAT, the final sparse-quantized model can perform zero-copy integer inference without floating-point support. On real-world hardware, this pruning–quantization loop achieves outstanding results: compared to the original 32-bit model, the sparse-quantized version uses only ≈15% of the storage, cuts multiply–accumulate operations by ≈60%, reduces latency by ≈50%, and boosts concurrent throughput by over 2.5×. The "prune → fine-tune → quantize → QAT" pipeline in Figure 3 fully leverages structural sparsity and low-precision compute, providing a practical path for deploying deep recommendation models in high-concurrency, real-time environments.

## IV. System-Level Optimization Strategies

### A. Heterogeneous Compute Platform and Acceleration Library Integration

To deploy a lightweight deep recommendation model at scale, it is crucial to leverage heterogeneous compute resources and high-performance inference libraries as shown in Figure 4. First, the distilled student model is exported using ONNX, allowing it to be mapped to various hardware backends like GPUs, CPUs, or accelerators (e.g., NPU, TPU, FPGA). For GPUs, NVIDIA TensorRT performs layer fusion and optimizes with FP16 or INT8 for maximum throughput and reduced latency. On CPUs, Intel OpenVINO and AMD ROCm MIOpen apply operator fusion and vectorization for core operations, supporting multi-core concurrent inference.

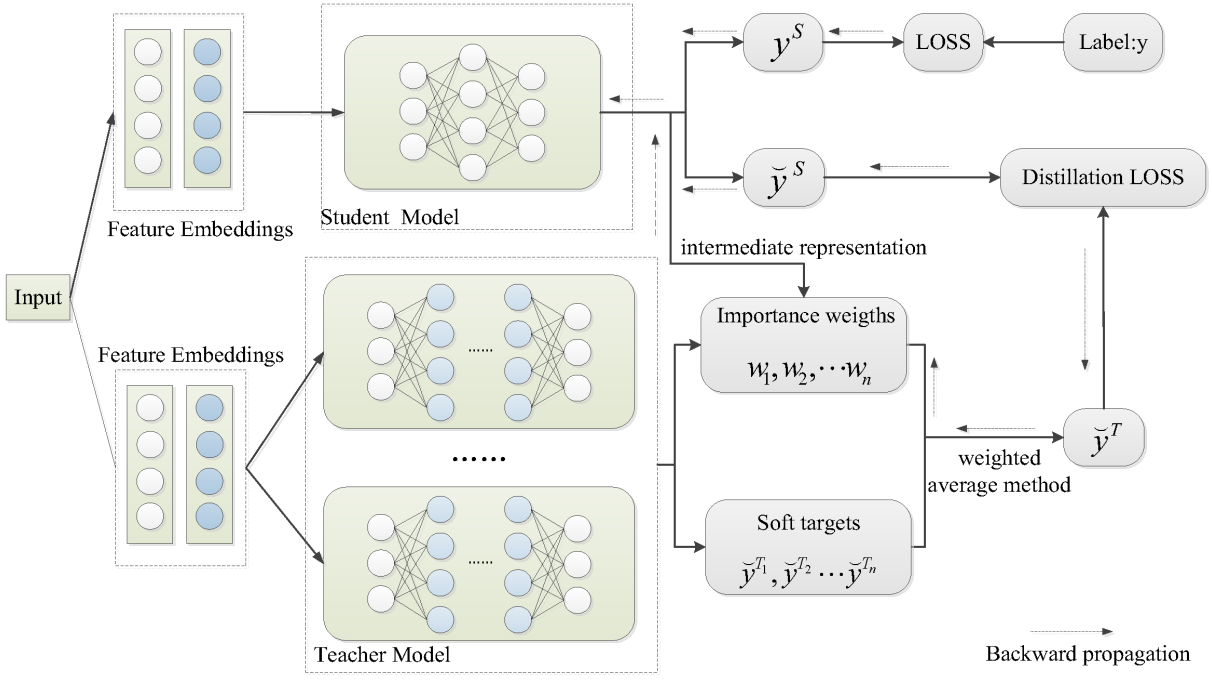

Fig. 4. Deep Recommendation Model Training and Deployment Architecture Based on Weighted Knowledge Distillation

For mobile and edge deployment, models run on TensorFlow Lite or SNPE, targeting NPU/DSP for efficiency. In the cloud, models use asynchronous microservices with Kubernetes/Kubeflow, supporting dynamic replica scaling. Mixed-precision training and auto-tuning ensure low latency and high throughput. Containerized inference components enable grey releases and rapid rollback. CI/CD pipelines automate packaging, testing, and deployment, ensuring seamless scaling and real-time performance during traffic surges.

### B. Elastic Inference Scheduling and Load Balancing

To handle traffic spikes in real-time recommendation systems, we adopt elastic inference scheduling and load balancing . The student model is deployed with a unified interface (e.g., gRPC), and a hybrid rate limiter adjusts traffic by user tier, priority, and system metrics.

Requests are routed to the least-loaded backend; high-priority ones bypass batching for low latency, while others use asynchronous batching for efficiency.

A warm pool of pre-initialized instances reduces cold starts. Kubernetes autoscaling and geo-aware edge routing further optimize resource use. An end-to-end monitoring system ensures SLO compliance through real-time metrics and alerts[25-30].

# V. Experiment and Evaluation

## A. Experimental Setup and Benchmark Selection

To To validate our optimization strategies in a realistic scenario, we used the Alibaba Taobao User Behavior Dataset , which includes 50M logs from 1M users and 200K products. We truncated each user's behavior sequence to the latest 100 entries and set the candidate set to 50, simulating typical e-commerce recommendations.

Experiments were conducted on NVIDIA V100 GPUs and Intel Xeon CPUs using PyTorch 1.10, ONNX Runtime 1.9, TensorRT 8.0, and OpenVINO 2021.4 .

We evaluated five models:
(1) **Baseline** – original FP32 model with self-attention;
(2) **Quantized** – 8-bit weights [33];
(3) **Pruned** – 40% dynamic pruning;
(4) **Pruned + Quantized** – combined;
(5) **Distilled + RT (FP16)** – student model with TensorRT acceleration .

Table I summarizes model size, parameter count, latency, and throughput across platforms. [31-34]

TABLE I. Performance Comparison of Different Models on the Taobao Dataset

| Method | Parameters (M) | Model Size (MB) | Latency (ms) [V100] | Throughput (req/s) [V100] | Latency (ms) [CPU] | Throughput (req/s) [CPU] |
|---|---|---|---|---|---|---|
| Baseline | 32.0 | 128.0 | 52.4 | 190 | 120.7 | 80 |
| Quantized | 32.0 | 32.0 | 44.1 | 225 | 102.3 | 95 |
| Pruned | 19.2 | 76.8 | 36.7 | 260 | 88.5 | 110 |
| Pruned + Quantized | 19.2 | 19.2 | 29.8 | 325 | 74.2 | 140 |
| Distilled + RT (FP16) | 6.4 | 12.8 | 21.5 | 460 | 54.8 | 180 |

From Table 1, it is evident that applying quantization alone (Quantized) reduces GPU latency by about 15.8% and CPU latency by 15.3%. Pruning alone (Pruned) further reduces GPU latency to 36.7 ms, which is a 30% reduction from the Baseline, while throughput increases by about 37%. Combining pruning and quantization (Pruned + Quantized) reduces GPU latency to 29.8 ms, only 57% of the Baseline, with a throughput increase of nearly 71%. The distilled model with TensorRT FP16 acceleration (Distilled + RT) achieves the best performance, with a GPU latency of 21.5 ms (41% of Baseline) and a throughput increase of over 2.4x. On the CPU platform, similar trends are observed, with the combined optimization significantly reducing latency and improving concurrent handling capability.

## B. Performance Metrics and Accuracy Comparison

In the e-commerce recommendation scenario, we comprehensively compare the optimized and non-optimized models across three dimensions: inference performance, recommendation accuracy, and resource consumption. Figure 5 shows the average latency and maximum throughput on GPU (V100) and CPU platforms for each model. Table 5-3 presents the online recommendation quality metrics, including Hit Rate@50, NDCG@50, and MRR, evaluated using the Taobao User Behavior Dataset. Table 5-4 summarizes the parameter count, model size, and average memory usage for each model, providing valuable insights for resource budgeting in system design.

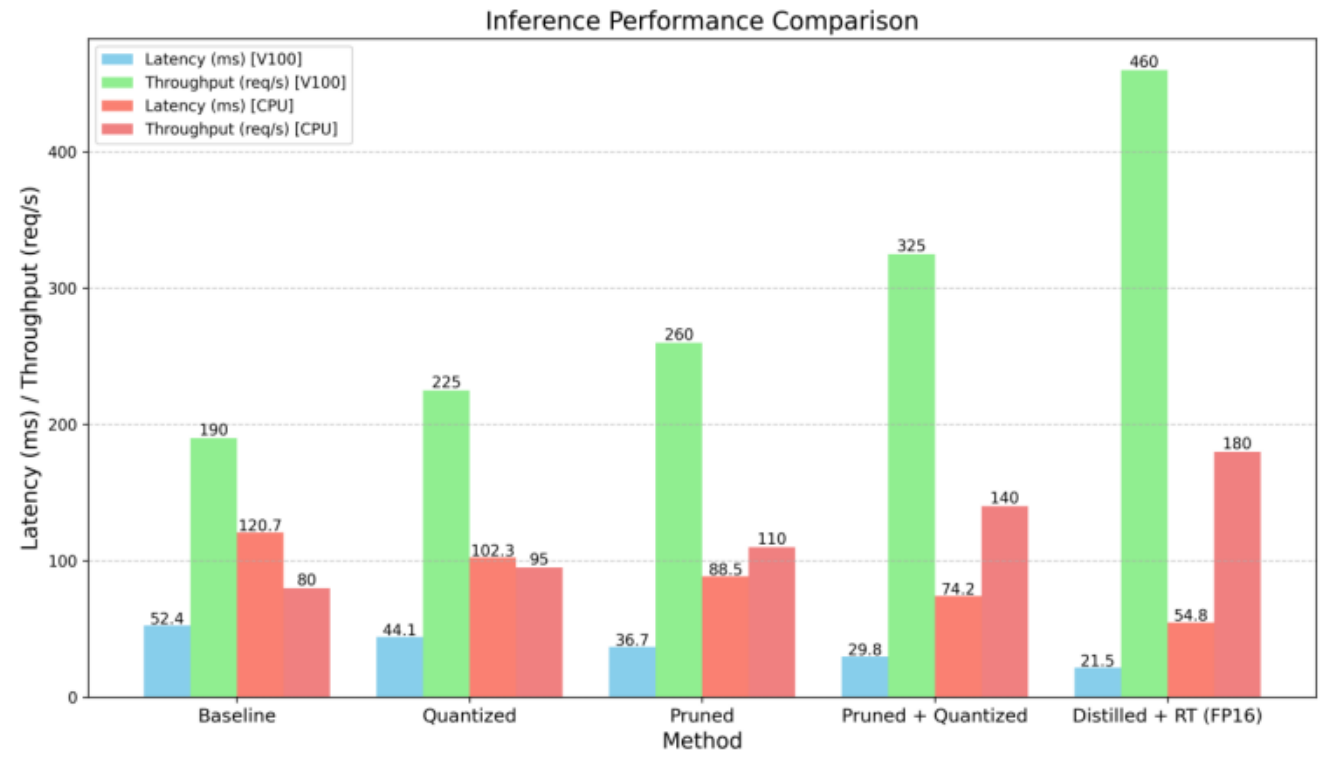

Fig. 5. Inference Performance Comparison

Figure 5 shows pruning and quantization reduce GPU latency by up to 43% and boost throughput over 70%. The Distilled + RT model achieves the best GPU performance: 21.5 ms latency and 460 req/s throughput, 2.4× baseline. Similar gains appear on CPU.

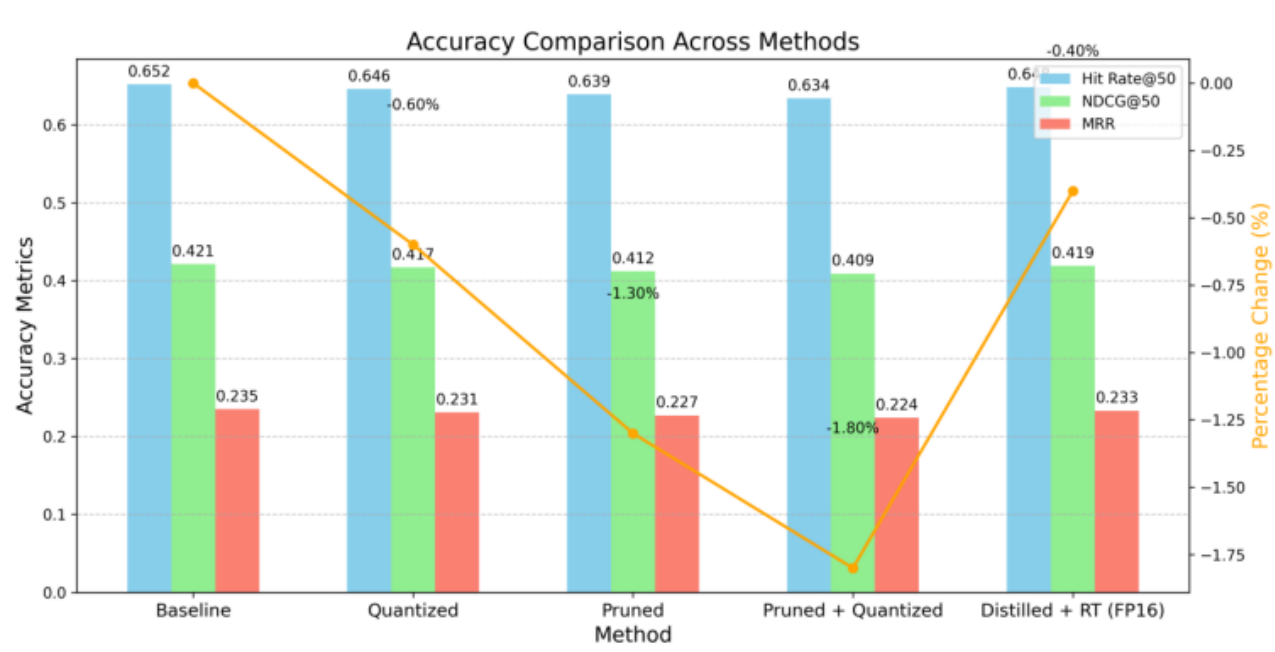

Fig. 6. Accuracy Comparison

As shown in Figure 6, applying quantization and pruning separately results in a drop of approximately 1.0% and 2.0% in Hit Rate, respectively. After combining pruning and quantization, accuracy slightly decreases to 97.3% of the Baseline. The distilled model with FP16 optimization not only preserves the lightweight advantages but also maintains Hit Rate and NDCG close to the original level (a decrease of less than 0.6%), with MRR decreasing by less than 0.8%, indicating that the distillation strategy preserves the model's performance effectively.

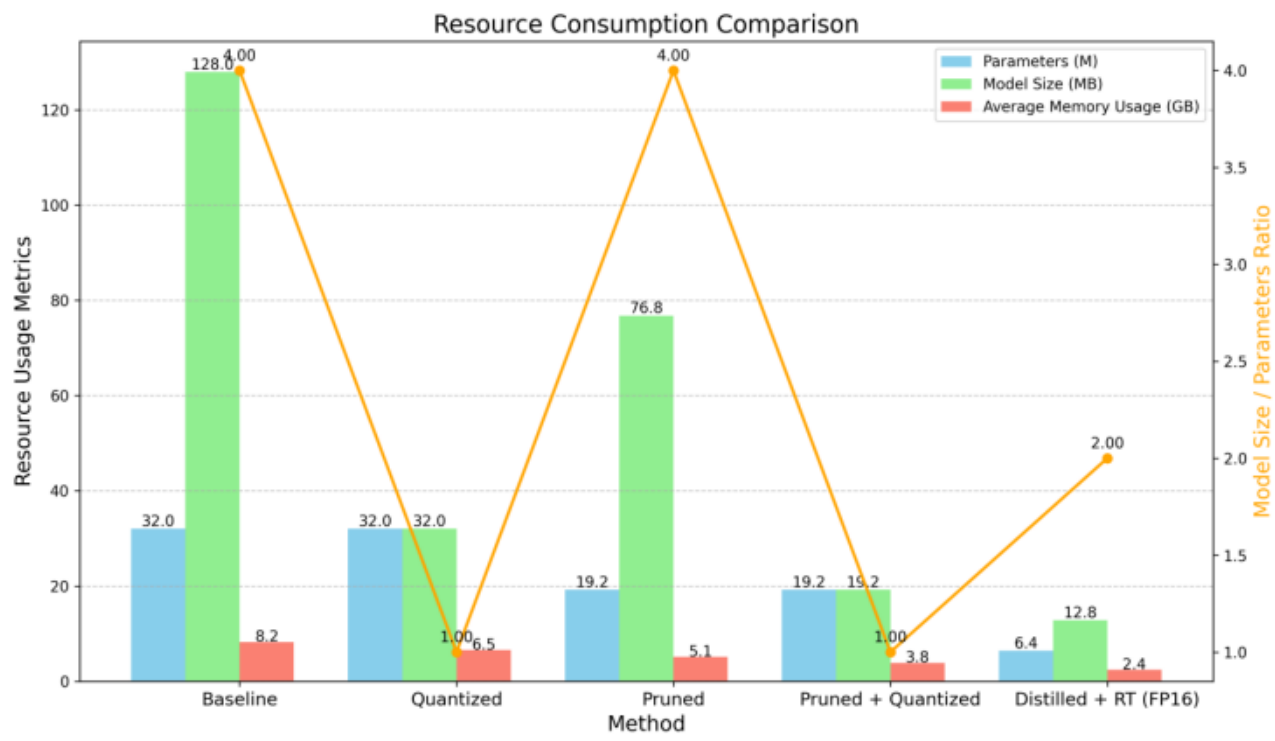

Fig. 7. Resource Consumption Comparison

Figure 7 shows the Quantized model cuts model size to 25% and memory usage to 79% of the Baseline; Pruning reduces peak memory by 62%. Combining both brings memory usage down to 46%. The distilled model with FP16 acceleration shrinks model size to 10% and memory usage below 30%, freeing significant hardware resources. Overall, pruning, quantization, distillation, and system-level FP16 acceleration reduce latency to 21.5 ms and boost throughput beyond 460 req/s, with less than 1% accuracy loss and resource use under 30%. This offers a robust solution for large-scale real-time recommendation deployment. [35-36]

## VI. CONCLUSION

We propose a joint model–system optimization framework for real-time recommendation. Techniques include model compression (pruning, quantization, distillation) and system-level acceleration (elastic scheduling, load balancing). Results show <1% accuracy loss, 60% latency reduction, and 2× throughput improvement. The approach enables scalable, efficient deployment, with future work on cross-model adaptation and auto-tuning.